\documentclass[12pt]{article}
\usepackage[english]{babel}
\usepackage[T1]{fontenc}
\usepackage{newtxtext,newtxmath}  
\usepackage{graphicx}
\usepackage{setspace}
\usepackage{bm}
\usepackage{cite}
\usepackage{amsmath}
\usepackage{amsfonts}
\usepackage{newtxtext,newtxmath}
\usepackage{algorithm}
\usepackage{algorithmic}
\usepackage{graphicx}
\usepackage{caption}
\usepackage{subcaption}
\usepackage{hyperref}
\usepackage{booktabs}
\usepackage{multirow}
\usepackage{enumitem}
\usepackage{adjustbox}
\usepackage{array}
\usepackage{titlesec}
\usepackage[left=1in, right=1in, top=1in, bottom=1in]{geometry}
\usepackage{indentfirst}
\usepackage{lipsum}
\titleformat{\section}[block]{\rmfamily\bfseries}{\thesection}{1em}{}

\titlespacing*{\section}{0pt}{0pt}{0pt}

\newcommand{\headingfont}{\rmfamily}

\begin{document}

	\begin{center}
		{\headingfont\fontsize{18}{22}\selectfont\textbf{Transmitter Identification via Volterra Series Based Radio Frequency Fingerprint}}
	\end{center}
	
	\begin{center}
		\headingfont\textbf{Rundong Jiang, Jun Hu, Zhiyuan Xie, Yunqi Song, Shiyou Xu}
		
		School of Electronics and Communication Engineering, Sun Yat-Sen University\\
	\end{center}
	
	\begin{center}
		\headingfont\textbf{\textit{Abstract}}
	\end{center}
	
	\noindent
	{The explosive proliferation of wireless communication devices has made secure network access increasingly challenging. 
Radio Frequency Fingerprinting (RFF), a physical-layer authentication technique that requires no cryptographic pairing and exhibits inherent resistance to cloning or spoofing, has emerged as a promising solution for device authentication in the Internet of Things (IoT) and unmanned systems. 
However, existing studies often lack a unified theoretical definition of RFF and effective feature extraction methodologies. 
Most existing approaches either rely on handcrafted features derived from specific signal attributes or directly employ neural networks for classification, resulting in limited generalization and poor interpretability.
Inspired by nonlinear system analysis, we model the transmitter as a black box whose internal behavior is expressed through its effect on the transmitted waveform. 
By interpreting the deviation between the ideal reference signal and the actual transmitted signal as a manifestation of hardware-induced distortions, we compose the received signal using the Volterra series and employ its kernels as comprehensive device features that capture both linear and nonlinear hardware characteristics. 
To address the high-dimensionality of higher-order Volterra kernels, we approximate them via wavelet decomposition and estimate their coefficients through a least-squares fitting process. 
The resulting wavelet coefficients serve as concise yet informative representations of hardware behavior and are subsequently classified using a complex-valued neural network. 
Experimental validation on a public LoRa dataset demonstrates state-of-the-art performance, achieving over 98\% accuracy in static channels and above 90\% accuracy under multipath and Doppler conditions. 
The proposed method not only enhances interpretability but also generalizes well across channel environments. 
The implementation is publicly available at: 
\url{https://github.com/thomas-smith123/RFFI}\\
\textit{This work has been submitted to the IEEE for possible publication. Copyright may be transferred without notice, after which this version may no longer be accessible.}
	}
	
	\noindent
	
	\textit{{\headingfont\textbf{Key Words:Radio frequency fingerprint, Volterra series, Wavelet decomposition}} }
	
	\setstretch{1.2}
	\section{Introduction}
	Wthe rapid evolution of wireless communication technologies and the pervasive deployment of the IoT, ensuring secure access to wireless networks has become a fundamental challenge. Traditional authentication mechanisms typically rely on cryptographic methods such as pre-shared keys or digital certificates. 
While these approaches offer high security, they require complex key management and distribution frameworks and remain vulnerable to various attacks, including man-in-the-middle and replay attacks. Consequently, a lightweight, reliable, and hardware-rooted authentication mechanism is essential for future wireless systems.

RFF is a hardware-based authentication technology that identifies devices through unique radio characteristics inherently induced by hardware imperfections. Such characteristics, including carrier frequency offset, amplitude distortion, modulation errors, and transient response anomalies, originate from inevitable manufacturing tolerances of analog components. 
As these imperfections are extremely difficult to replicate, RFF provides a robust identity signature for each device. Combined with advanced signal analysis and machine learning techniques, RFF can effectively support device identification, authentication, and anti-counterfeiting in IoT and wireless networks.

An RFF encapsulates the cumulative effects of linear and nonlinear distortions introduced by multiple hardware stages during signal transmission. As the signal propagates through mixers, filters, and power amplifiers, each component imposes its distinct transformation, and the composite effect reflects the device's intrinsic hardware behavior \cite{wangModelBasedRFFingerprint2025}. 
However, despite significant research efforts, there remains no universally accepted definition or extraction framework for RFF. 
Current mainstream RFF approaches fall into two categories: feature-based methods that rely on handcrafted features derived from domain expertise, and data-driven methods that use deep neural networks to automatically learn features from raw signals. 
The former suffers from limited generalization across modulation schemes, while the latter, although powerful, lacks interpretability and physical insight.

To provide a more principled understanding, we adopt a hardware-system perspective and define the RFF as the difference between the ideal transmitted signal and the actual emitted signal \cite{jiangGeneralizedRadioFrequency2024}. 
This definition explicitly associates the fingerprint with the hardware-induced deviations of the transmitter, making it physically interpretable and mathematically tractable.

\begin{figure}[htbp]
    \centering
    \includegraphics[width=8cm]{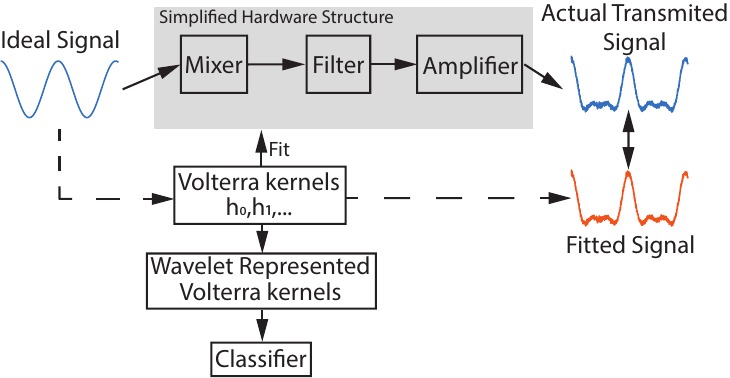}
    \caption{Overall Framework}
    \label{fig:overall}
\end{figure}
The Volterra series, a classical framework for nonlinear system modeling, provides a natural mathematical basis for this definition. 
It represents the input–output relationship of a system as a functional expansion, where the first-order kernel captures linear behavior and higher-order kernels characterize nonlinearities and memory effects \cite{centurelliMethodsModelComplexity2022,nypwipwyPowerAmplifierBehavioral2022}. 
By composing the transmitted signal using the Volterra series via idea transmitted signal, we can represent hardware behavior through kernel functions that encode both amplitude and phase distortions caused by nonlinear circuit elements. 
These kernels, therefore, serve as holistic RFF representations.

To manage the computational complexity associated with high-order kernels, we truncate the Volterra expansion to the second order, which retains most nonlinear characteristics while remaining computationally feasible. Moreover, we approximate Volterra kernels using wavelet scaling functions, providing compact representations that preserve both temporal and spectral localization. This wavelet-based approximation substantially reduces parameter dimensionality while enhancing robustness to noise and sampling variations.

Given the widespread use of complex I/Q signals in modern wireless systems, the extracted kernel coefficients are inherently complex-valued. 
Consequently, we design a complex-valued neural network classifier to exploit both amplitude and phase information during feature learning. 
This network architecture demonstrates superior performance in capturing the complex correlations inherent in RFF.

Our main contributions are summarized as follows:
\begin{itemize}
    \item We introduce the Volterra series framework into RFF analysis, interpreting the transmitter as a nonlinear system whose kernel parameters serve as physically meaningful fingerprint features.
    \item We propose a wavelet-based approximation for Volterra kernels, significantly reducing model dimensionality and improving parameter stability while retaining essential nonlinear characteristics.
    \item Experimental validation on public LoRa datasets demonstrates superior accuracy and generalization compared with state-of-the-art RFF methods.
\end{itemize}

\section{Related Works}
\label{sec:related_work}
In this section, we mainly review RF system analysis methods and RFF extraction methods.
\subsection{RF System Analysis Methods}
Methods for analyzing RF systems can generally be divided into two major categories: linear and nonlinear analysis. 
Linear characterization remains fundamental for evaluating amplifier gain, impedance matching, isolation, and stability, with the classical S-parameter framework serving as the analytical cornerstone. 
However, the transformations that occur when a signal passes through hardware include not only linear transformations but also nonlinear transformations.
Therefore, accurate hardware modeling must incorporate nonlinear transformations in addition to linear ones.

Traditional nonlinear modeling approaches often rely on polynomial representations such as the Taylor series, which are effective for weakly nonlinear systems without substantial memory effects~\cite{ermolovaSpectralAnalysisNonlinear2001}. 
Yet, in wideband and high-power devices, nonlinearities are often strong and exhibit memory-dependent behavior. 
In such cases, Taylor-series-based models fail to accurately capture large-signal phenomena such as compression, saturation, and intermodulation distortion. 

To address these limitations, the Volterra series has become a cornerstone in nonlinear system modeling, especially in characterizing RF power amplifiers~\cite{zhuDecomposedVectorRotationBased2015,bocciarelliHighAccuracyLowCostGeneralized2023}. 
The Volterra framework systematically describes the input–output relationship of nonlinear systems with memory, offering a unified representation that encompasses both harmonic generation and intermodulation effects. 
Each Volterra kernel corresponds to a specific order of nonlinearity, and its frequency-domain formulation provides valuable insights into harmonic distortion and spectral regrowth. 
Recent research has focused on efficient kernel estimation algorithms and kernel simplification techniques to balance accuracy and computational efficiency~\cite{chenLowComplexityMoving2022}.

Beyond Volterra-based models, behavioral modeling frameworks such as the X-parameter method have been developed as an extension of S-parameters to large-signal regimes~\cite{yangXparametersModelingBased2025}. 
X-parameters linearize the system behavior around a specific large-signal operating point and capture the nonlinear relationships between fundamental and harmonic components. 
Although X-parameter models can predict device performance under complex modulation schemes with good accuracy, they are limited by measurement complexity and poor extrapolation capability outside calibrated regions.

More recently, data-driven modeling paradigms have emerged as powerful alternatives to analytical methods. 
Deep neural networks (DNNs) can directly learn complex nonlinear mappings between input and output signals from measured data~\cite{xuChebyshevPolynomialLSTMModel2022,dosreisLowComplexityLSTMNNBased2022}. 
Unlike traditional physical models, DNNs inherently account for nonlinearity, memory effects, and even component-level variations. 
These models have demonstrated performance on par with, or even superior to, classical Volterra-based and X-parameter approaches, providing new tools for system-level analysis and hardware characterization.

\subsection{RF Fingerprint Extraction Methods}
In RFF-based individual emitter classification, how to obtain the RFF is crucial. 
Generally, most RFF extraction methods focus on detailed hardware impairments. 
There are two mainstream approaches: one is manually designed based on experience, and the other is using deep learning to automatically extract features. 
Experience-based methods differ from deep learning-based methods because the former has lower intelligence and relies on expert experience for feature selection. 
It is well known that RFF originate from hardware. 
Hardware-level defects of the device are reflected in the final output signal. 
Although the RF system analysis methods mentioned in the previous section cover many analysis methods for RF systems, current RFF extraction methods still mostly focus on extracting certain specific signal features. 
Defects in hardware devices such as power amplifiers \cite{wangWirelessPhysicalLayerIdentification2016a}, clocks, mixers lead to signal distortion, clock jitter \cite{pengDesignHybridRF2019b}, I/Q imbalance \cite{zhuoRadioFrequencyFingerprint2017}, phase noise, DC offset \cite{xingRadioFrequencyFingerprint2018a}, etc. These features can be extracted from statistical features of the signal \cite{bertonciniWaveletFingerprintingRadioFrequency2012b}, energy features \cite{koseRFFingerprintingIoT2019}, frequency stability of the signal \cite{shenRadioFrequencyFingerprint2021}, ambiguity function \cite{xinRadioFrequencyFingerprint2025}, bispectrum \cite{nieUAVDetectionIdentification2021b}, etc. Additionally, statistical indicators of time-domain signals are also used as RFF, such as kurtosis, skewness, and variance \cite{linWirelessDeviceIdentification2020,miContentindependentMethodLFM2022,tuResearchInternetThings2019}. Experience-based RFF extraction methods require a large amount of prior information and complex manual analysis and calculation. However, different types of signals often have different prior information, so these methods are often limited by poor adaptability.

Deep learning mainly involves data-driven feature extraction. Based on learning from a large number of samples, generalized and dataset-specific feature representations can be obtained, improving the efficiency and accuracy of dataset representation. The extracted abstract features are more robust and have better generalization ability. In many RFF feature extraction applications, the main method is to directly classify the signal \cite{jagannathEmbeddingAssistedAttentionalDeep2023} or to classify after preprocessing. For example, image-based methods include converting the signal into time-frequency images through Wigner-Ville distribution, short-time Fourier transform, etc. \cite{liuRadioFrequencyFingerprint2023}, differential constellation traces \cite{wangDesignNoiseRobust2024,zhangFineGrainedRadioFrequency2023}, etc. Then, neural networks are used to extract features from the images, and finally, a classifier is used for classification.

Regarding the use of deep learning methods, increasingly new network architectures are being applied in this field. Not only the most classic convolutional neural networks \cite{shenDeepLearningPowered2023}, but also autoencoders \cite{yaoEfficientRFFExtraction2023}, Vision Transformer \cite{huiCrossAttentionTransformerChannelRobust2025}, LSTM \cite{jiangRadioFrequencyFingerprint2024}, attention mechanisms \cite{zhangDataandKnowledgeDualDrivenRadio2023}, and other deep learning models or improved structures are used for signal feature extraction, classification, and recognition. The focus of these methods is on designing and improving new network structures to achieve higher accuracy on datasets. Although they can automatically learn features, they lack interpretability, making it difficult to understand the basis for model decisions. The incompleteness of RFF modeling leads to an unclear understanding of RFF characteristics and interference impacts, resulting in insufficient analysis of the effectiveness and limitations of existing interference mitigation methods and feature extractor designs.

\section{Our Proposed Methodology}

In this section, we present the theoretical foundation and practical implementation of the proposed RFF extraction and identification framework. 
Our approach models the transmitter as a nonlinear dynamic system using the Volterra series, which provides a unified and physically interpretable formulation for hardware-induced distortions. 
The Volterra kernels are then approximated using wavelet bases, and their coefficients are estimated via least-squares fitting. Finally, the fitted coefficients are input to a complex-valued neural network for classification.


\subsection{Transmission Model}
\begin{figure}[htbp]
    \centering
    \includegraphics[width=8cm]{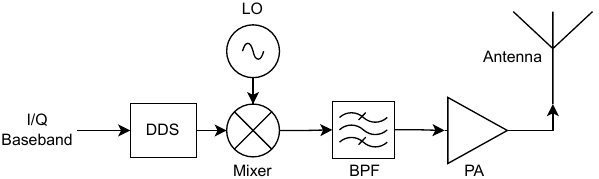}
    \caption{RF System Transmission Model}
    \label{fig:tx_system_model}
\end{figure}
A simplified RF transmitter structure is shown in Fig. \ref{fig:tx_system_model}. 
The signal source generates the analog baseband signal through a Direct Digital Synthesizer (DDS). 
Then, a Bandpass Filter (BPF) filters out possible image frequencies and higher-order intermodulation components. Finally, a Power Amplifier (PA) amplifies the filtered signal to increase its power, and the signal is transmitted through the antenna.

However, in this seemingly simple transmission chain, each component affects the wireless signal, leaving unique characteristics in the frequency and time domains.
First, in the process from the signal source to the analog baseband signal, the DDS generates the analog I/Q waveform. Hardware defects can cause frequency offset, phase noise, and I/Q imbalance. During the mixing process, the Local Oscillator signal source (typically a Phase-Locked Loop) further introduces phase noise and frequency offset. 
Secondly, the nonlinear characteristics of the mixer itself lead to the generation of image frequencies and higher-order intermodulation products. Therefore, a BPF is used to filter the signal and attenuate spurious signals. 
However, the non-ideal in-band flatness of the BPF can also cause amplitude distortion and phase distortion in the signal. 
Finally, the nonlinear characteristics of the PA cause amplitude distortion and harmonic generation. 
In summary, the finally emitted signal is the result of a series of linear and nonlinear distortions. Even a single feature like frequency offset is often the result of the combined action of multiple components.

From the perspective of the received signal, the received signal is affected not only by the transmitter's hardware characteristics but also by the channel and the receiver's own hardware characteristics \cite{yanRRFRobustRadiometric2022}. 
In this paper, we only discuss the case of a single receiver. 
Channel effects generally include dispersion, attenuation, clutter, multipath, Doppler shift, noise, etc. 
Without additional processing and compensation for channel effects, channel effects, like receiver hardware, introduce additional distortion, affecting recognition accuracy. 
Strong channel effects can even mask the transmitter's hardware characteristics, leading to a significant decrease in recognition accuracy. This paper also only considers transmitter factors.

\subsection{Proposed Method}
During the process from signal generation at the information source to final transmission and reception, the signal undergoes processing by multiple components. 
Each stage introduces certain hardware impairments. 
Breaking away from the traditional thinking of feature extraction and neural network structure design, we analyze from the perspective of the transmitter system. 
Then, due to hardware impairments, there is a difference between the signal actually emitted by the transmitter and the ideal signal. 
We therefore model the transmitter as a black box that imposes both linear and nonlinear distortions on the ideal signal, producing deviations that characterize the device-specific behavior.
We use the behavior of this black box to define the RFF, i.e., the difference between the ideal signal and the actual emitted signal.


Analyzing from the transmission perspective, for nonlinear distortion of signals, we use Volterra series as analysis method. Its basic idea is to represent the system's output as a polynomial function of the input \cite{bieNonlinearBehavioralAnalysis2024,wangOpenSetRFFingerprint2024,hanDecomposedVectorCombinationBased2024}. 
Specifically, the Volterra series expresses the system's output as the sum of multiple integral terms of the input signal, each term containing the input signal and its past values raised to a certain power. 
We then take the parameters of the Volterra series as RFF features, aiming to enhance the interpretability and applicability of the method.

Therefore, after the ideal signal $u(t)$ passes through the hardware circuit, the signal can be expressed as:
\begin{equation}\label{eq:volterra_signal}
    \begin{aligned}
        y(t)=&h_{0}+\sum_{n=1}^{\infty}\int_{-\infty}^{\infty}\cdots\int_{-\infty}^{\infty}h_{n}(\tau_{1},\tau_{2},\cdots,\tau_{n})\cdot \\
        &u(t-\tau_{1})u(t-\tau_{2})\cdots u(t-\tau_{n})\,d\tau_{1}d\tau_{2}\cdots d\tau_{n}
    \end{aligned}
\end{equation}
where $h_n(\tau_1, \tau_2, \cdots, \tau_n)$ represents the $n$-th order Volterra kernel function of the system, describing the system's linear or nonlinear response to the input signal. 
$h_0$ represents the introduced constant offset, $h_1$ is the first-order kernel function representing the system's linear response, $h_2$ is the second-order kernel function representing the system's second-order nonlinear response, and so on. 
That is, the output of the signal is the superposition of the linear response and nonlinear responses of different orders of the input signal. The combined result of this series of hardware impairments is the RFF of the transmitting device, reflected in the series of kernel functions in Eq. \ref{eq:volterra_signal}.

Writing Eq.\ref{eq:volterra_signal} in discrete form:
\begin{equation}\label{eq:discrete_volterra}
    \begin{aligned}
        y[n]= & h_{0}+\sum_{\tau}h_{1}(\tau)u[n-\tau]+\\
        & \sum_{\tau_1}\sum_{\tau_2}h_{2}(\tau_{1},\tau_{2})u[n-\tau_{1}]u[n-\tau_{2}]+\cdots
    \end{aligned}
\end{equation}
For different devices that transmit the signal with same content, i.e., $u$ is the same, the difference in the signal emitted by the device depends only on the parameters of the Volterra kernels. 
Therefore, the parameters of the Volterra kernels can be considered as parameters that integrate all hardware impairments and can be used as RFF for individual emitter identification. 
According to practical situations, higher-order components gradually decrease in energy as the order increases. 
Therefore, for nonlinear system analysis, the Volterra series is usually truncated to the second order. 
Furthermore, for higher-order scenarios, the computational complexity increases significantly and is difficult to compute with current technology. 
Therefore, the method in this paper truncates the Volterra series to the second order.

For the first-order Volterra term:
\begin{equation}\label{eq:first_order_term}
    y_{1}[n]=\sum_\tau h_{1}(\tau)u[n-\tau]
\end{equation}
The first-order kernel function $h_1(\tau)$ can be approximated at scale $j$ using scaling functions as:
\begin{equation}
    h_{1}(\tau)~\approx~\sum_{j,k}\alpha_{j,k}\,\Phi_{j,k}(\tau)
\end{equation}
where $\Phi_{j,k}(\tau) = 2^{j/2} \Phi(2^j \tau - k)$ is the wavelet basis function, $\Phi$ is the wavelet scaling function, and $\alpha_{j,k}$ is the expansion coefficient. In the following text, we denote $j, k$ as $m$, indicating that the scale and shift of the $m$-th wavelet basis are $j$ and $k$, respectively.

For the second-order kernel:
\begin{equation}\label{eq:second_order_term}
    y_{2}[n]=\sum_{\tau_{1}} \sum_{\tau_{2}} h_{2}(\tau_{1},\tau_{2})\,u[n-\tau_{1}]u[n-\tau_{2}]
\end{equation}
Expand the second-order Volterra kernel using wavelets:
\begin{equation}
    \begin{array}{l}{{h_{2}(\tau_{1},{\tau}_{2})\approx\sum_{j_1,k_1,j_2,k_2}\beta_{j_1,k_1,j_2,k_2}\Phi_{j_1,k_1}(\tau_{1})\Phi_{j_2,k_2}({\tau}_{2})}}\end{array}
\end{equation}
Similar to the treatment of $j, k$ earlier, we denote $j_1, k_1$ as $m$ and $j_2, k_2$ as $n$. The second-order term can be written as:
\begin{equation}
    y_{2}[n]\approx\sum_{m,n}\beta_{m,n}\sum_{\tau_{1}} \sum_{\tau_{2}}\Phi_{m}(\tau_{1})\Phi_{n}(\tau_{2})\,u[n-\tau_{1}]u[n-\tau_{2}]
\end{equation}
We then estimate the $\alpha$ and $\beta$ parameters based on the actual signal transmitted by the transmitter and the ideal transmitted signal.

\subsection{Parameter Estimation}
In most RFF-based authentication scenarios, the transmitted signal's structure and parameters are known a priori.
Therefore, by adjusting the $\alpha$ and $\beta$ parameters above, the ideal transmitted signal can be synthesized by the second-order Volterra series to get a signal as close as possible to the actual transmitted signal \cite{peretsModelingNonlinearSignal2025}. 

For the first-order term, define:
\begin{equation}
	\mathbf{Z}^{(1)}=\left[ \begin{matrix}
		Z_{1}[0]&		...&		Z_{L_1}[0]\\
		\vdots&		&		\vdots\\
		Z_{1}[N-1]&		...&		Z_{L_1}[N-1]\\
	\end{matrix} \right] 
\end{equation}
where $\mathbf{Z}^{(1)} \in \mathbb{C}^{N \times L_1}$, $N$ is the number of signal points, for the convenience of writing, our N starts from 1. 
$L_1$ is the number of wavelet scaling functions, and $Z^{(1)}_m[n] = \sum_{\tau} \Phi_m(\tau)u[n-\tau]$. 
Represent $\mathbf{Z}^{(1)}$ as $\mathbf{Z}^{(1)} = [\mathbf{Z}^{(1)}_1, \mathbf{Z}^{(1)}_2, \cdots, \mathbf{Z}^{(1)}_{L_1}]$. 
Then, for the second-order features:
\begin{equation}
    \begin{aligned}
        &\mathbf{Z}^{(2)}=[\,{Z}^{(1)}_{i}\,\odot\,{Z}^{(1)}_{j}\,]_{(i,j)\in\Omega}=\\
        &\left[ 
        \begin{matrix}
            Z_{1}^{(1)}[1]Z_{1}^{(1)}[1]&		Z_{1}^{(1)}[1]Z_{2}^{(1)}[1]&		...&		Z_{L_2}^{(1)}[1]Z_{L_2}^{(1)}[1]\\
            Z_{1}^{(1)}[2]Z_{1}^{(1)}[2]&		Z_{1}^{(1)}[2]Z_{2}^{(1)}[2]&		...&		Z_{L_2}^{(1)}[2]Z_{L_2}^{(1)}[2]\\
            \vdots&		\ddots&		\ddots&		\vdots\\
            Z_{1}^{(1)}[N]Z_{1}^{(1)}[N]&		Z_{1}^{(1)}[N]Z_{2}^{(1)}[N]&		...&		Z_{L_2}^{(1)}[N]Z_{L_2}^{(1)}[N]\\
        \end{matrix} \right]
    \end{aligned}
\end{equation}
where $\odot$ denotes the Hadamard product, $\Omega = \{(i,j) | 1 \leq i \leq j \leq L_1\}$, $\mathbf{Z}^{(2)} \in \mathbb{C}^{N \times L_2}$, and $L_2 = \frac{L_1(L_1 + 1)}{2}$.

Then, define
\begin{equation}
	{\boldsymbol {\theta}_{\mathrm{ext}}}=\left[ h_0\;\vert\;\boldsymbol{\alpha}\;\vert\;\boldsymbol{\beta} \right]^T
\end{equation}
\begin{equation}
    {\bf Z}_{\mathrm{ext}}=\left[{\bf1}\;\vert\;{\bf Z}^{(1)}\;\vert\;{\bf Z}^{(2)}\right]
\end{equation}
where, $\boldsymbol{\alpha} = [\alpha_1, ..., \alpha_{L_1}]^T$; $\boldsymbol{\beta} = [\beta_{11}, \beta_{12}, ..., \beta_{L_1 L_1}]^T$. 
Then Eq. \ref{eq:discrete_volterra} can be written as:
\begin{equation}
    \mathbf{y} = \mathbf{Z}_{\text{ext}} \boldsymbol{\theta}_{\mathrm{ext}} + \boldsymbol{\epsilon}
\end{equation}
where $\mathbf{y}$ is the actual received signal and $\boldsymbol{\epsilon}$ is the truncation error. By setting $\min \| \mathbf{y} - \mathbf{Z}_{\mathrm{ext}} \boldsymbol{\theta} \|_2^2$ as the optimization objective, the RFF $\boldsymbol{\theta}_{\mathrm{ext}}$ for the current signal can be obtained. The optimization objective is as shown in Eq. \ref{eq:optimization_target}:
\begin{equation}\label{eq:optimization_target}
    {\hat{\boldsymbol{\theta}}}_{\mathrm{ext}}={\underset{\boldsymbol{\theta}_{\mathrm{ext}}}{\operatorname{arg\,min}}}\,\|\mathbf{y}-\mathbf{Z}_{\mathrm{ext}}\boldsymbol{\theta}_{\mathrm{ext}}\|_{2}^{2}+\lambda\|\boldsymbol{\theta}_{\mathrm{ext}}\|_{2}^{2}
\end{equation}
where $\lambda$ is the regularization parameter. 
The analytical solution to Eq. \ref{eq:optimization_target} is:
\begin{equation}
	{\hat{\boldsymbol{\theta}}}_{\mathrm{ext}}=\left(\mathbf{Z}_{\mathrm{ext}}^{H}\mathbf{Z}_{\mathrm{ext}}+\lambda\mathbf{I}\right)^{-1}\mathbf{Z}_{\mathrm{ext}}^{H}\mathbf{y}
\end{equation}
where $\mathbf{Z}^H$ is the Hermitian transpose.

In this paper, we use the least squares method for parameter estimation.

\subsection{Classifier Design}
\begin{figure}[htbp]
    \centering
    \includegraphics[width=7.5cm]{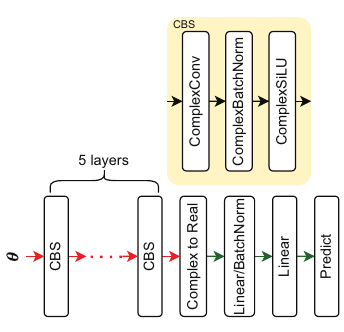}
    \caption{Structure of Classifier}
    \label{fig:Classifier_structure}
\end{figure}
After fitting the Volterra kernels, we use the kernel parameters $\hat{\boldsymbol{\theta}}_{\text{ext}}$ as the RFF and use a classifier for classification. Since the signal parameters used in the fitting process are complex-valued, the fitted parameters are also complex-valued. Therefore, we use a simple complex-valued neural network to classify the parameters. 
The structure of the used complex network is shown in Fig. \ref{fig:Classifier_structure}. 
In the figure, the red arrow represents the complex flow, and the green arrow represents the real flow. 
First, we use 5 layers of CBS blocks connected in series. 
Here, the CBS layer represents a series connection of Complex 1D Convolution, Complex 1D Batch Normalization layer, and Complex SiLU activation function. The Complex SiLU function is implemented by applying the SiLU activation function separately to the real and imaginary parts. For complex-to-real conversion, we use Eq. \ref{eq:complex2real} for implementation \cite{jiangCoarsetoFineSpectrumMonitoring2025}:
\begin{equation}\label{eq:complex2real}
    \mathbf{Real}(\mathbf{Z}) = \frac{\mathbf{X}+\mathbf{Y}}{\sqrt{2}}
\end{equation}
where $Z = X + jY$, and $X$ and $Y$ are both real numbers.

Finally, two fully connected layers are used to map the results into predicted labels.
During the classifier training process, the cross entropy function is used as the loss function so that ${\hat{\boldsymbol{\theta}}}_{\mathrm{ext}}$ can be mapped to the predicted label with the highest accuracy, as shown in Eq. \ref{eq:celoss}.
\begin{equation}\label{eq:celoss}
    L = -\sum_{i=1}^{C} y_i \log(p_i)
\end{equation}
where, $C$ is the number of categories, $y_i$ is the true label, and $p_i$ is the predicted probability.

\section{Experimental Validation}
\begin{figure}[htbp]
	\centering
	\subfloat[Fitting result on dataset without multipath and Doppler frequency offset]{\includegraphics[width=0.45\textwidth]{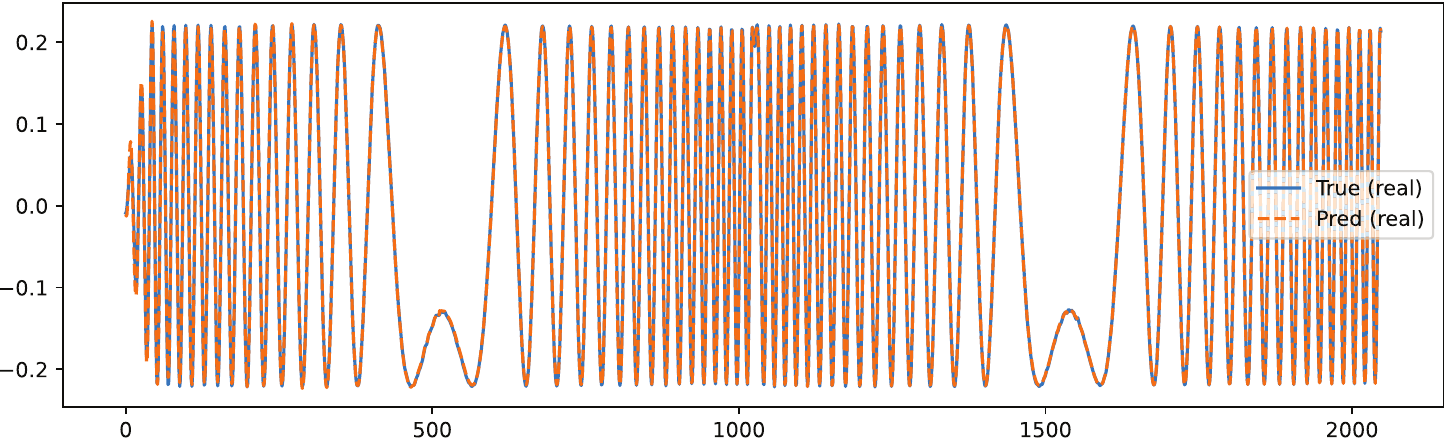}}

	\subfloat[Fitting result on dataset with multipath and Doppler frequency offset]{\includegraphics[width=0.45\textwidth]{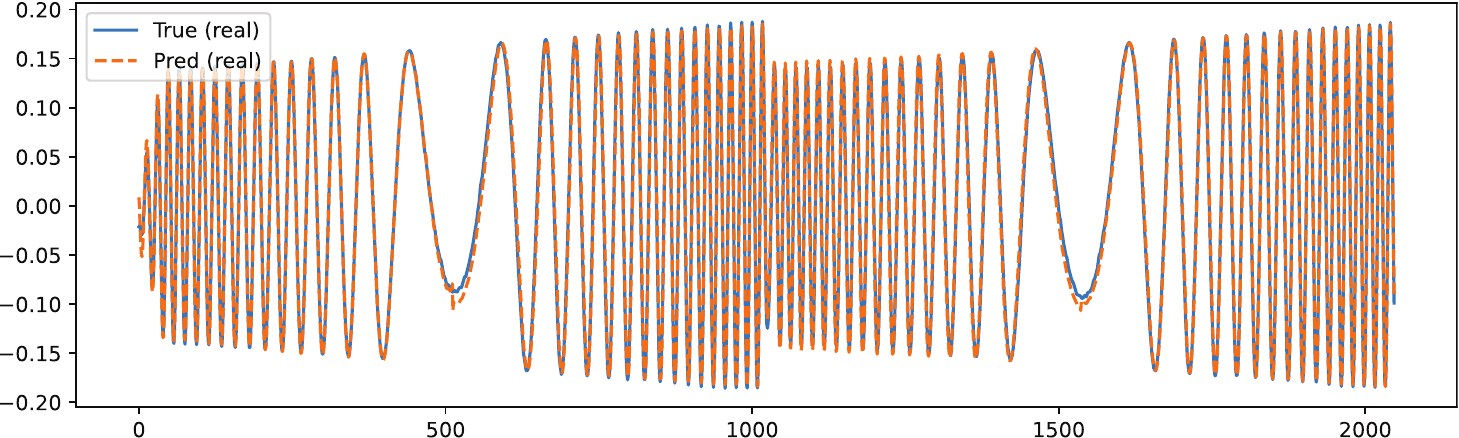}}
	\caption{Comparison of Fitting Results} 
    \label{fig:fit_situation}
\end{figure}
We conducted experimental validation using a public LoRa dataset. 
We used the dataset from \cite{shenScalableChannelRobustRadio2022} and performed validation on three datasets: without multipath and Doppler frequency offset, with multipath only, and with both multipath and Doppler frequency offset. 
The dataset without multipath and Doppler frequency offset contains signals collected from 30 different LoRa devices under static conditions. 
Each device has 500 signals, each with 8192 sampling points, including all parts of the 8-symbol LoRa protocol preamble. 
The datasets with multipath only and with multipath and Doppler frequency offset also have 30 devices, but each device has 1000 signals.
We first conducted experiments and validation on the dataset without multipath and Doppler frequency offset. 
For the known-parameter LoRa signals, we first reconstructed the ideal LoRa signal preamble part based on the parameters of the sampled data, as $u(t)$ in Eq. \ref{eq:first_order_term} and Eq. \ref{eq:second_order_term}. 
We first fitted the signals of the LoRa dataset using the aforementioned multi-wavelet combined with Volterra series method to obtain the Volterra kernel parameters for each signal. 
We used 3 wavelet basis functions. The dimensionality of the extracted features was 910. We then used a classifier composed of a complex convolutional network and fully connected layers for classification. The complex convolution in Fig. \ref{fig:Classifier_structure} is 1D convolution with a kernel size of 9, stride of 2, and padding of 4.

\subsection{Experimental Results}

We first verified the fitting effect of reconstructing the original signal using the Volterra kernels decomposed by wavelets with the ideal signal. 
The fitting results are shown in Fig. \ref{fig:fit_situation}. 
The figure shows that the fitting effect is generally good, verifying the effectiveness of our method. 
However, for signals with multipath and Doppler frequency offset, there are relatively large fitting errors in some places.

\begin{figure*}[htbp]
	\centering
	\subfloat[PCA result on features on dataset without multipath and Doppler frequency offset]{\includegraphics[width=0.3\textwidth]{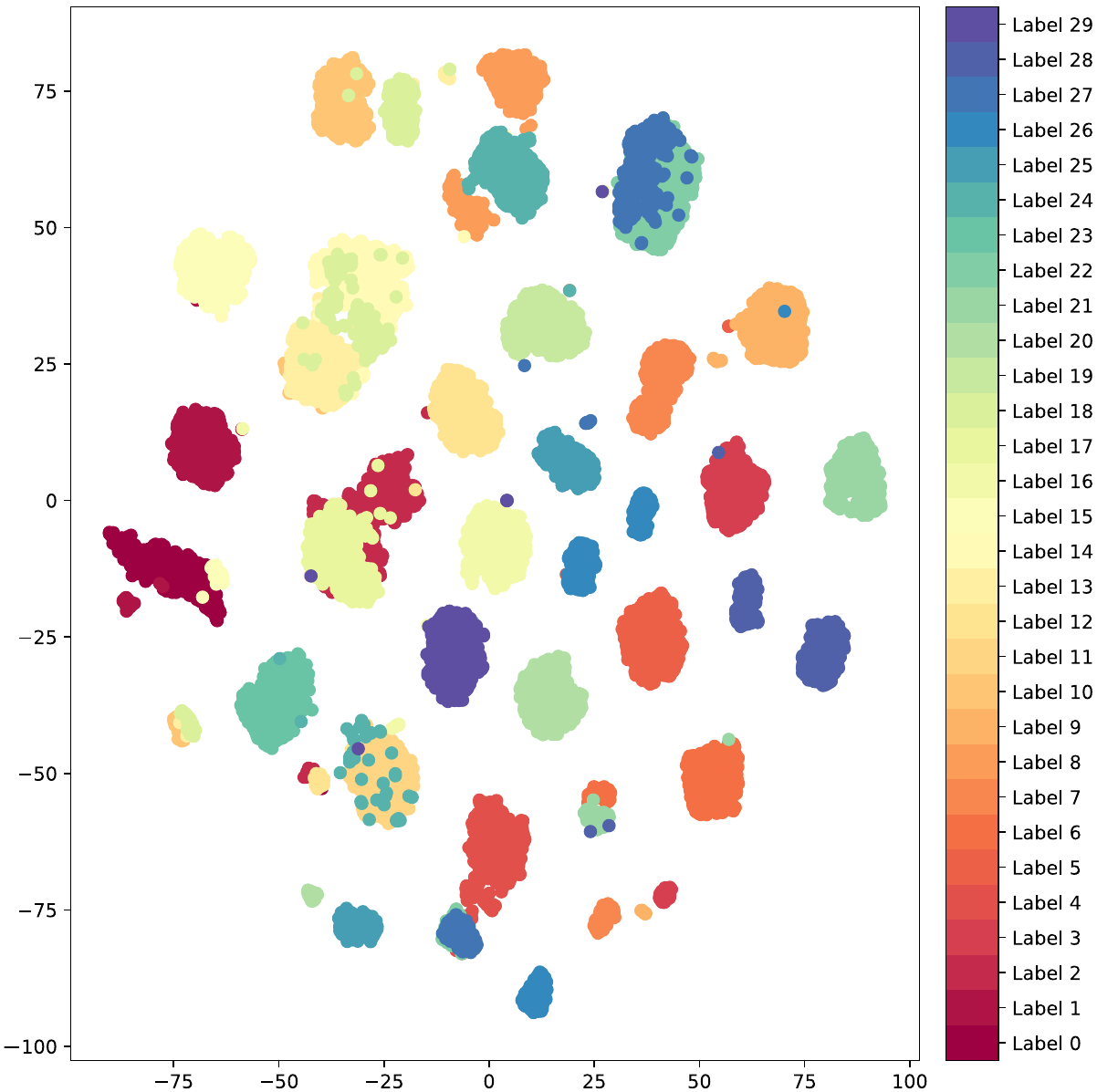}}
	\subfloat[PCA result on features on dataset with multipath only]{\includegraphics[width=0.3\textwidth]{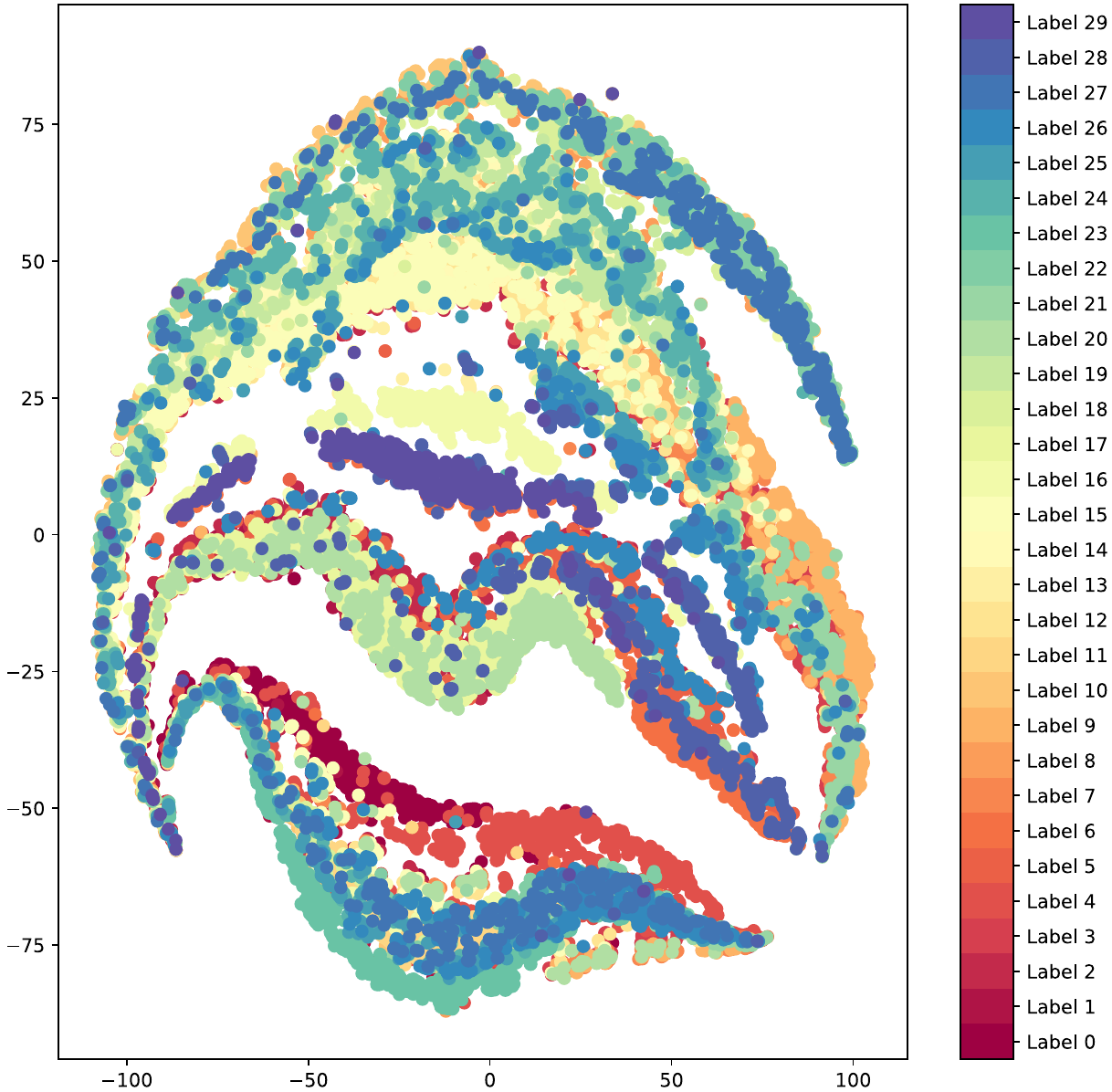}}
    \subfloat[PCA result on features on dataset with multipath and Doppler frequency offset]{\includegraphics[width=0.3\textwidth]{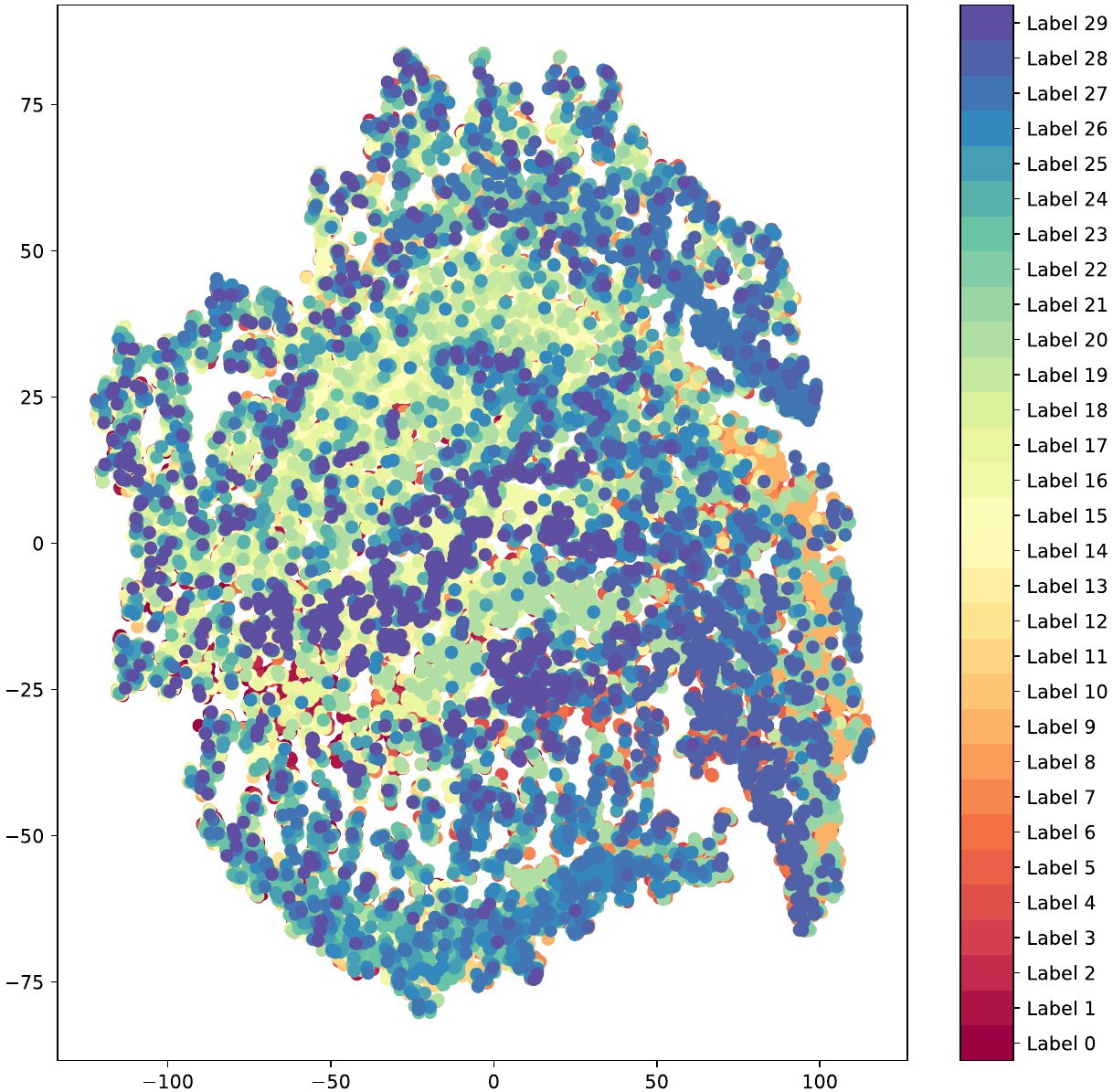}}
	\caption{PCA Result of Extracted Features on Different Datasets} 
    \label{fig:pca_result}
\end{figure*}
Subsequently, we visualized the features of the fitted kernel parameters. 
The results are shown in Fig. \ref{fig:pca_result}. 
It can be seen from the figure that without any multipath and Doppler frequency offset, the PCA dimensionality reduction features of the wavelet coefficients obtained by directly fitting the Volterra series using wavelets already have high separability. 
Then, as multipath effects and Doppler frequency offset are added, the separability of the features gradually decreases, until they appear completely mixed. 
Therefore, to further improve classification accuracy, we use a complex convolutional neural network as a classifier to further classify the features.

\begin{figure}[htbp]
    \centering
    \includegraphics[width=7.5cm]{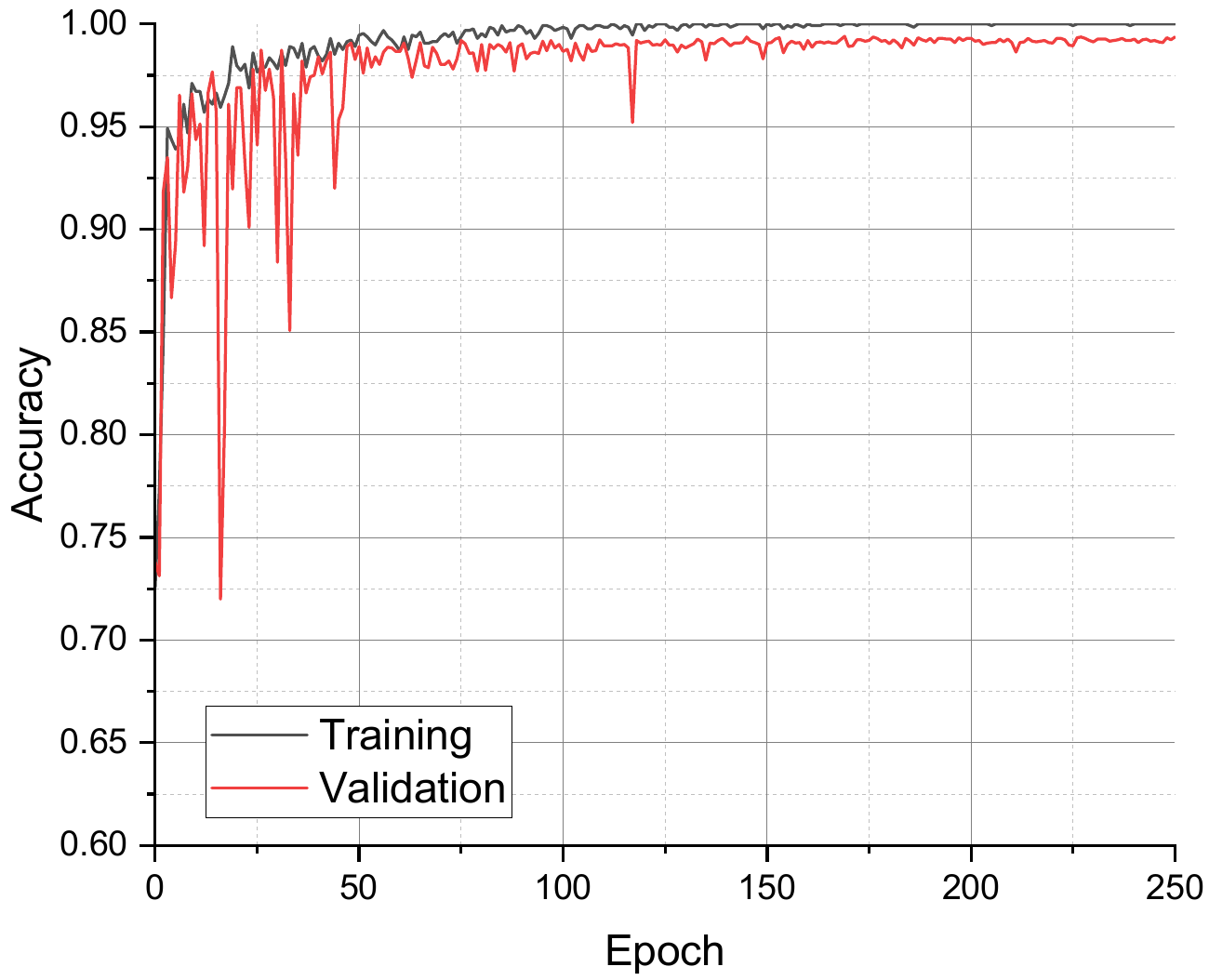}
    \caption{Training and Validation Situation}
    \label{fig:Training_and_Validation}
\end{figure}

Using the features from the dataset without frequency offset and multipath, we used 80\% of the samples as the training set and the remaining 20\% as the validation set. The training process is shown in Fig. \ref{fig:Training_and_Validation}. We conducted experiments with the number of $j$ being 3. 
Our experimental results are shown in Fig. \ref{fig:Training_and_Validation}. In the figure, the black line is the classification accuracy on the training set, and the red line is the classification accuracy on the validation set. During training, the accuracy on the training set exhibited large fluctuations in the early stages and then gradually stabilized. The validation set followed a similar trend, reaching nearly 99\% accuracy at the end of training. 

\begin{figure}[htbp]
	\centering
	\subfloat[Confusion matrix on Training dataset]{\label{subfig:cm_train}\includegraphics[width=0.45\textwidth]{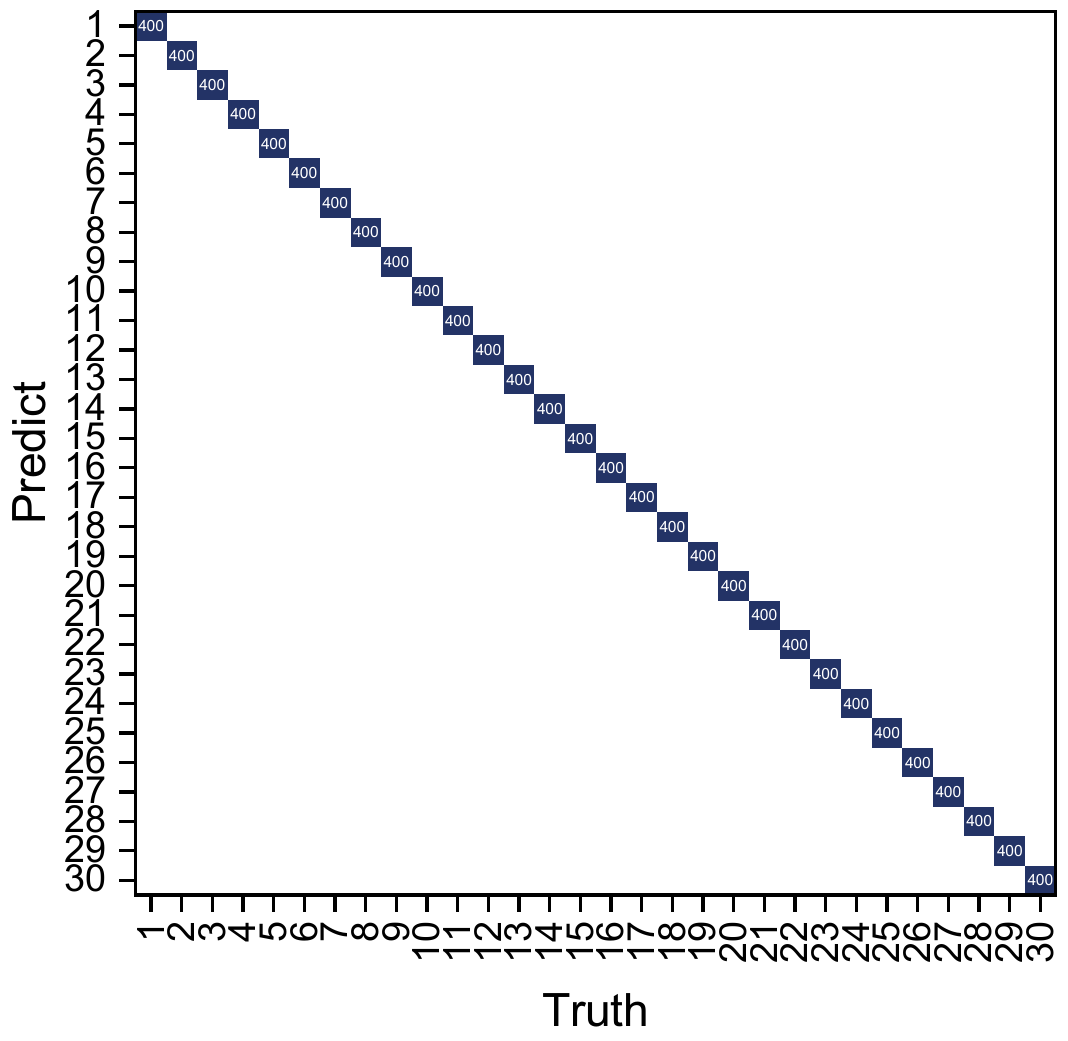}}

	\subfloat[Confusion matrix on Validation dataset]{\label{subfig:cm_valid}\includegraphics[width=0.45\textwidth]{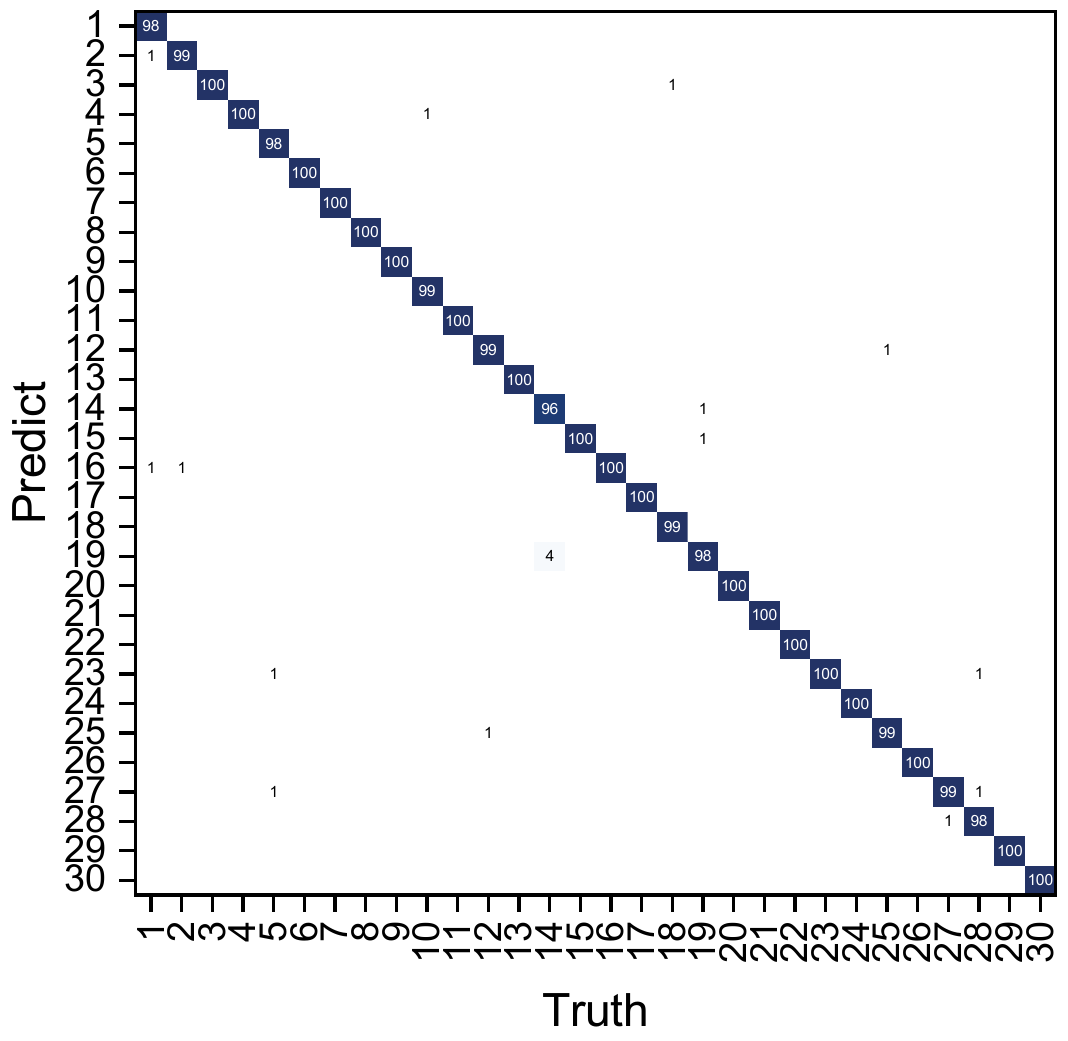}}
	\caption{Confusion Matrix on Training and Validation dataset} 
    \label{fig:Confusion_Matrix_valid}
\end{figure}

Having high accuracy demonstrates the effectiveness of our method. From the training results, our method did not show severe overfitting. We tested the classifier after training. The confusion matrices on the training set and validation set are shown in Fig. \ref{subfig:cm_train} and Fig. \ref{subfig:cm_valid}, respectively. 

In the figures, the numbers on the confusion matrix represent the number of actual samples classified into that label. It can be seen that for 30 categories of samples, our classification accuracy on the training set reached 100\%. On the validation set, except for one class at 96\%, the accuracy of other categories was no less than 98\%.

\begin{figure}
    \centering
    \includegraphics[width=7.5cm]{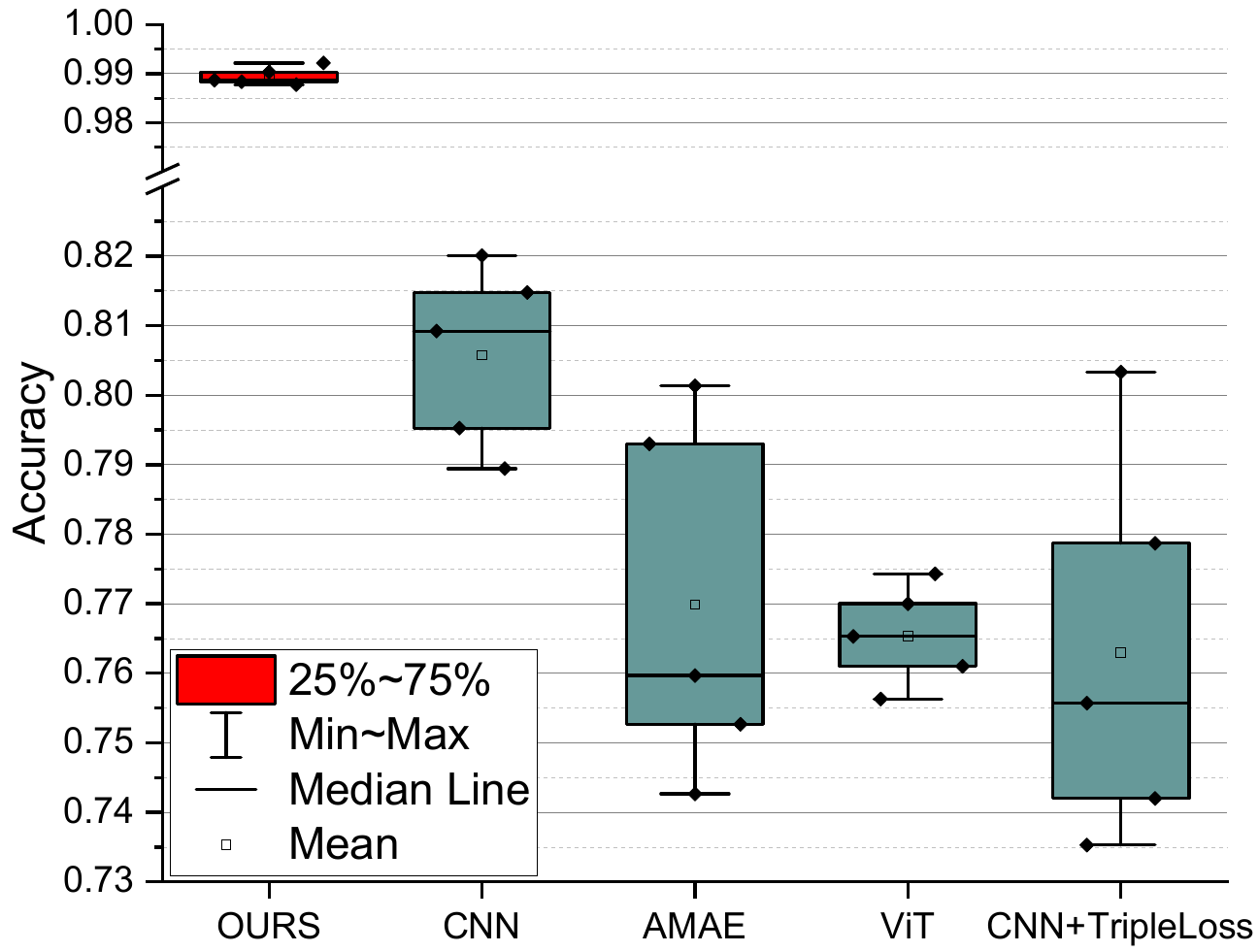}
    \caption{Compare With Other Methods on 5-fold Cross Validation}
    \label{fig:5cross_compare}
\end{figure}

To verify the generalization of our method, we performed 5-fold cross-validation on the same dataset and compared the results with other methods, as shown in Fig. \ref{fig:5cross_compare}. The box plots from the second left are for methods using CNN \cite{shenDeepLearningPowered2023}, Symmetric Autoencoder \cite{yaoEfficientRFFExtraction2023}, ViT, and CNN combined with Triple Loss \cite{shenScalableChannelRobustRadio2022}. In the figure, the red box plot is our result. The position of the black hollow square represents the mean accuracy. The horizontal line in the box is the median. The upper and lower edges of the box represent the upper and lower quartiles, respectively. The horizontal lines outside the box represent the maximum and minimum values. The black diamonds represent the positions of the sample points in the current statistical sample. In the figure, our mean accuracy is 98.94\%, the highest accuracy is 99.22\%, and the lowest accuracy is 98.77\%.

\begin{figure}
    \centering
    \includegraphics[width=7.5cm]{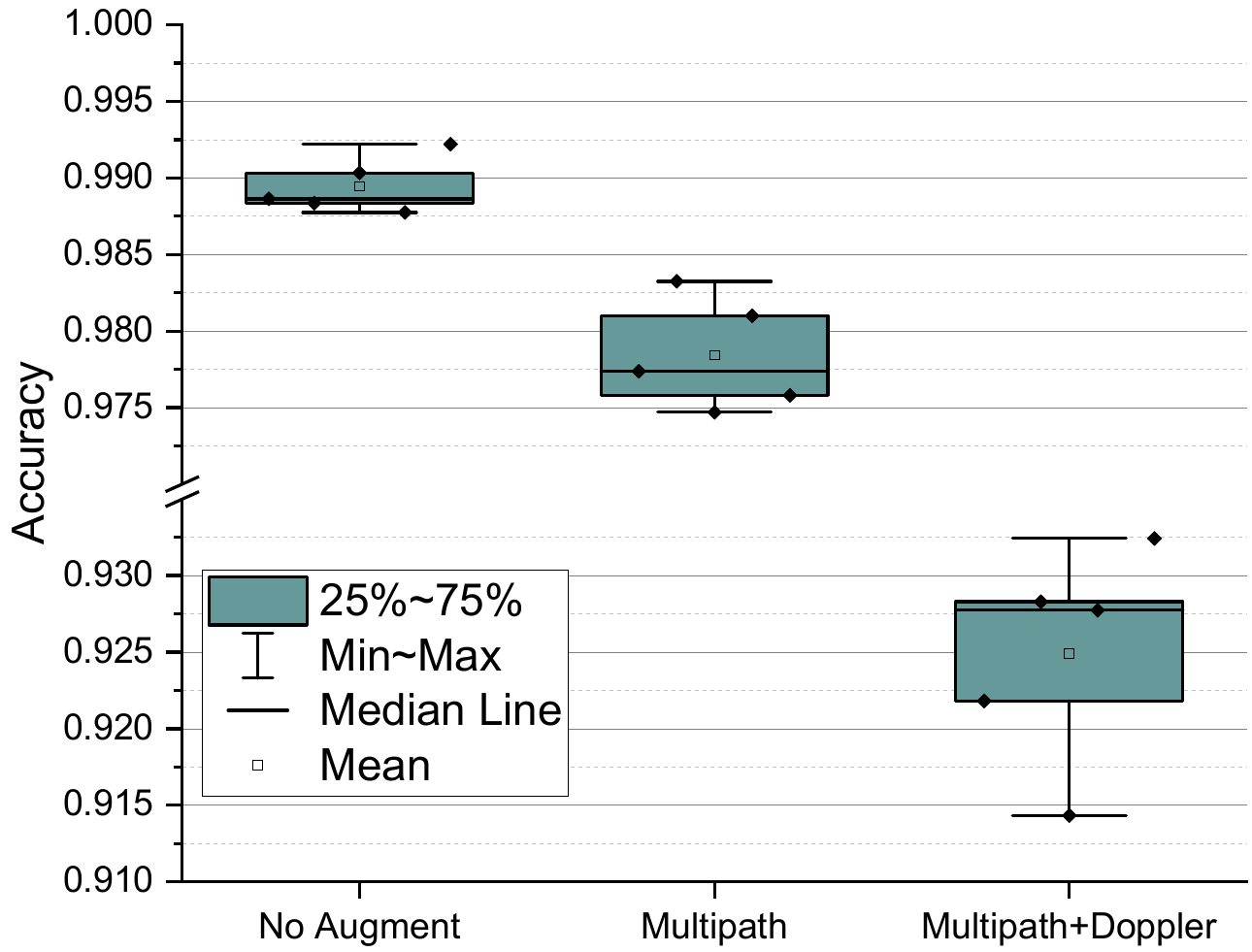}
    \caption{Compare With Other Channel Effects on 5-fold Cross Validation}
    \label{fig:5cross_compare_aug}
\end{figure}

Subsequently, we also performed 5-fold cross-validation experiments on the augmented dataset with multipath and frequency shift. The experimental results are shown in Fig. \ref{fig:5cross_compare_aug}.
From the results, it can be seen that as some channel effects and interference are added to the signal, the accuracy on the validation set gradually decreases. For augmented samples with only multipath, the mean classification accuracy is greater than 97.75\%, and the lowest accuracy is slightly below 97.5\%. For samples with both Doppler frequency offset and multipath, the accuracy further decreases, with a mean accuracy of only about 92.5\%, but the lowest accuracy is greater than 91.75\%. We analyze that the multipath effect causes the signal itself to overlap in the time dimension, which also causes some overlap in the time-domain features, thereby slightly affecting classification accuracy. As for the introduction of Doppler frequency offset, it can be explained from a hardware perspective as the hardware characteristics exhibited by the same hardware at different frequency points being different, causing a greater impact on accuracy.

\section{Conclusion}
This paper approaches the problem from the perspective of the transmitter's hardware composition, analyzing how the transmitter affects the emitted signal, and defines the RFF as the difference between the ideal signal and the actual transmitted signal. We treat the entire transmitter as a black box and introduce a method to decompose linear and higher-order nonlinear signal components using Volterra series within the RFF system. Given known input and output signals, the Volterra series is used to model the behavior of the black box, with the resulting Volterra kernels serving as RFF features for classification. When the Volterra kernel parameters are numerous and difficult to estimate, we apply wavelet decomposition to the kernels and use the resulting wavelet coefficients as features, ultimately employing a neural network for classification.

Based on the results of neural network training, our method shows potential for further improvement. Nonetheless, on the current dataset, it achieves higher accuracy compared to existing approaches applied to the same dataset. Theoretically, for digitally modulated signals, our method is less sensitive to the modulation scheme than other methods. However, it has certain limitations, including the requirement for knowledge of the signal modulation scheme and the ability to reconstruct the ideal signal. Furthermore, classification accuracy degrades in the presence of multipath effects and Doppler frequency offset. In future work, we plan to extend the method’s applicability to additional signal types, such as WiFi, and to further investigate feature extraction and classification under complex channel conditions, addressing the limitations in multipath scenarios mentioned above.

	\bibliographystyle{IEEEtran}
	\bibliography{Transmitter_Identification_via_Volterra_Series_Based_Radio_Frequency_Fingerprint}
	
\end{document}